%% file: ms.tex
\documentclass[12pt,preprint,usenatbib]{emulateapj}

\newcommand{\change}[1]{{#1}}

\newcommand{\teff}{$T_{\mathrm{eff}}$}
\newcommand{\muhz}{$\mu$Hz}
\newcommand{\numax}{$\nu_{\mathrm{max}}$}
\newcommand{\dnu}{$\Delta\nu$}
\newcommand{\astec}{\textsc{ASTEC}}
\newcommand{\adipls}{\textsc{ADIPLS}}

\shorttitle{Radius determination of solar-type stars}
\shortauthors{Stello et al.}

\begin{document}



\title{Radius determination of solar-type stars using asteroseismology: \\
    What to expect from the Kepler mission}



\author{Dennis Stello\altaffilmark{1,2}, 
  William J. Chaplin\altaffilmark{3},
  Hans Bruntt\altaffilmark{1},
  Orlagh L. Creevey\altaffilmark{4}, 
  Antonio Garc\'{\i}a-Hern\'andez\altaffilmark{5},
  Mario J. P. F. G. Monteiro\altaffilmark{6,7},
  Andr\'es Moya\altaffilmark{5},
  Pierre-Olivier Quirion\altaffilmark{2}, 
  Sergio G. Sousa\altaffilmark{6,7},
  Juan-Carlos Su\'arez\altaffilmark{5}, 
  Thierry Appourchaux\altaffilmark{8},
  Torben Arentoft\altaffilmark{2},
  Jerome Ballot\altaffilmark{9,10},
  Timothy~R.~Bedding\altaffilmark{1}, 
  J{\o}rgen~Christensen-Dalsgaard\altaffilmark{2},
  Yvonne~Elsworth\altaffilmark{3}, 
  Stephen T. Fletcher\altaffilmark{11},
  Rafael A. Garc\'ia\altaffilmark{12}, 
  G{\"u}nter~Houdek\altaffilmark{13,14},
  Sebastian J. Jim\'enez-Reyes\altaffilmark{4},
  Hans~Kjeldsen\altaffilmark{2},
  Roger~New\altaffilmark{10}, 
  Clara R\'egulo\altaffilmark{4,15},
  David Salabert\altaffilmark{4,16},
  Thierry Toutain\altaffilmark{3}} 

\altaffiltext{1}{Sydney Institute for Astronomy (SIfA), School of Physics, University of Sydney, NSW 2006,
  Australia; stello@physics.usyd.edu.au.}
\altaffiltext{2}{Danish AsteroSeismology Centre (DASC), Department of Physics and Astronomy, Aarhus University, 8000 Aarhus C, Denmark}
\altaffiltext{3}{School of Physics and Astronomy, University of Birmingham,
  Edgbaston, Birmingham, B15 2TT, UK}
\altaffiltext{4}{Instituto de Astrof\'isica de Canarias, E-38200, La
  Laguna, Tenerife, Spain}
\altaffiltext{5}{Instituto de Astrof\'isica de Andaluc\'ia, CSIC, CP3004,
  Granada, Spain}
\altaffiltext{6}{Centro de Astrof\'{\i}sica da Universidade do Porto, Rua das
  Estrelas, 4150-762 Porto, Portugal}
\altaffiltext{7}{Departamento de Matem\'atica Aplicada, Faculdade de 
  Ci\^encias da Universidade do Porto, Portugal}
\altaffiltext{8}{Institut d' Astrophysique Spatiale (IAS), UMR8617, Batiment 121,
  F-91405, Orsay Cedex, France}
\altaffiltext{9}{Max-Planck-Institut f\"ur Astrophysik,
  Karl-Schwarzschild-Str. 1, Postfach 1317, 85741, Garching, Germany}
\altaffiltext{10}{Laboratoire d~RAstrophysique de Toulouse-Tarbes,
  Universit de Toulouse, CNRS, 14 avenue Edouard Belin, F-31400 Toulouse, France}
\altaffiltext{11}{Faculty of Arts, Computing, Engineering and Sciences,
  Sheffield Hallam University, Sheffield S1 1WB, UK}
\altaffiltext{12}{Laboratoire AIM, CEA/DSM-CNRS-Universit\'e Paris Diderot;
  CEA, IRFU, SAp, centre de Saclay, F-91191, Gif-sur-Yvette, France}
\altaffiltext{13}{Institute of Astronomy, University of Vienna, A-1180
  Vienna, Austria} 
\altaffiltext{14}{Institute of Astronomy, University of Cambridge,
  Cambridge CB3 0HA, UK}
\altaffiltext{15}{Departamento de Astrof\'isica, Universidad de La Laguna,
  La Laguna, 38206, Tenerife, Spain}
\altaffiltext{16}{National Solar Observatory, 950 North Cherry Avenue,
  Tucson, AZ 85719, USA}

\clearpage

\begin{abstract}
{\change
For distant stars, as observed by the NASA Kepler satellite, parallax
information is currently of fairly low quality and is not complete. This
limits the precision with which the absolute sizes of the stars and their
potential transiting planets can be determined by traditional methods. 
Asteroseismology will be used to aid the radius 
determination of stars observed during NASA's Kepler mission. 
We report on the recent asteroFLAG hare-and-hounds Exercise\#2, 
where a group of `hares' simulated data of F-K main-sequence stars that 
a group of `hounds' sought to analyze, aimed at determining the stellar
radii. We investigated stars in the range
$9<V<15$, both with and without parallaxes. We further test
different uncertainties in \teff, and compare results with and without
using asteroseismic constraints.
Based on the asteroseismic large frequency spacing, obtained from
simulations of 4-year time series data from the Kepler mission, we
demonstrate that the stellar radii can be correctly and precisely
determined, when combined with traditional stellar parameters from the
Kepler Input Catalogue.  
The radii found by the various methods used by each independent hound
generally agree with the true values of the artificial stars to within 3\%,
when the large frequency spacing is used. This is 5--10 times better than
the results where seismology is not applied.
These results give strong confidence that radius estimation can be
performed to better than $3\%$ for solar-like stars using automatic
pipeline reduction. Even when the stellar distance and luminosity are
unknown we can obtain the same level of agreement.
Given the uncertainties used for this exercise we find that
the input $\log g$ and parallax do not help to constrain the radius, and that
\teff~and metallicity are the only parameters we need in addition to
the large frequency spacing. It is the uncertainty in the metallicity that
dominates the uncertainty in the radius.
}
\end{abstract}

\keywords{stars: fundamental parameters --- stars: oscillations --- stars: interiors}

\clearpage

\section{Introduction} 
With the recent successful launches of the CoRoT \citep{Baglin06} and Kepler
missions \citep{Borucki08} we have entered a new era with strong 
synergy between the 
fields of transiting exoplanets and stellar oscillations (asteroseismology).
This synergy exists because of the common requirements for long uninterrupted
high-precision time-series photometry, enabling the same data to be used
for both purposes, providing complementary information.
The investigation of
stellar oscillations, and especially solar-like oscillations, provides a
unique tool to probe the interiors of stars.
In particular,
asteroseismology can provide an independent radius estimate of a
planet-hosting star, which can then be used to constrain the size of its  
transiting planet(s) \citep{Dalsgaard07,Stello07a,Kjeldsen08}. 

The new space missions 
provide the first opportunity to measure solar-like oscillations in a large
number of F, G and K main-sequence stars, which have photometric amplitudes of only
a few parts per million. These oscillations are global p modes, excited
by near-surface convection.
The frequency spectra (or p-mode spectra) of the oscillations show a
characteristic spacing called the large frequency spacing, \dnu, between
modes of successive radial order $n$. This spacing scales as the square
root of the mean stellar density and can therefore be used to
constrain the stellar radius with very high precision.
Investigating this potential for the Kepler mission
has been carried out in the framework of the asteroFLAG collaboration, whose
aim is to develop and test robust tools for 
analyzing asteroseismic data on solar-like stars.  
AsteroFLAG includes 
members of the Kepler
Asteroseismic Science Consortium \citep[KASC; ][]{Dalsgaard07}, and of
the CoRoT asteroseismology team \citep{Appourchaux08}. 
Much of the work within asteroFLAG is founded on a series of hare-and-hound
exercises, where a group of `hares' generates artificial data that 
the `hounds' seek to analyze. 

The first part of the asteroFLAG investigation (Exercise\#1) reported
by \citet{Chaplin08} involved 
extracting the large frequency spacing from artificial Kepler data.   
In this paper we build on those results to test  
how reliably we can obtain the stellar radius. We assume the availability
of standard catalogue data (e.g. \teff, $\log g$, $V$, and parallax), which
will come from the Kepler Input Catalogue \citep{Brown05} or other sources. 

Of the more than 100,000 stars to be observed by the Kepler
satellite, most will be observed at low cadence (30 min), but a few hundred
targets will be observed in a high-cadence mode (1
min). The observing mode for a given target can be changed from low to high
cadence if there are indications that the target shows transit-like
events, which will benefit both the transit measurement as well as the
supporting asteroseismic investigation.
While low cadence is sufficient to sample solar-like 
oscillations in evolved subgiants and red giant stars that have periods of
hours to days, in this paper we focus 
on the prime targets of the Kepler mission -- the main-sequence stars. These
have oscillation periods of a few minutes to several tens of minutes,
hence requiring data in the high cadence mode for a successful application
of asteroseismology.  

\S~\ref{exercise1} will provide a summary of the asteroFLAG
hare-and-hounds Exercise\#1, which feeds as input to Exercise\#2
reported in \S~\ref{observations}. {\change The discussion in
  \S~\ref{observations} includes the results}. The detailed descriptions of
the methods {\change applied} by each hound are given in
\S~\ref{stello}--\ref{trialerror}, followed by a general discussion in
\S~\ref{lessons} on {\change how our results depend 
  on the input parameters, and we highlight possible systematic
  errors}. Finally, we give the conclusions in \S~\ref{conclusions}. 

\section{Exercise\#1 results}
\label{exercise1}
\begin{figure}
\includegraphics{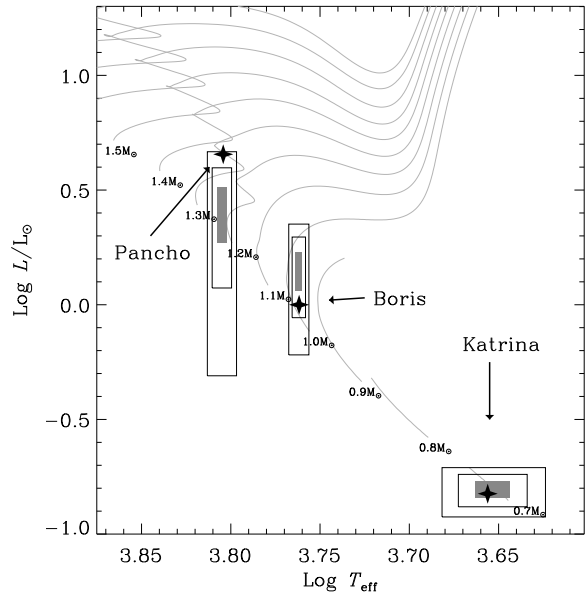}
\caption{H-R diagram showing evolution tracks (gray lines) and the true 
  position of the three artificial stars (star symbols), {\change which were not
  known by the hounds. The $1\sigma$, $2\sigma$, and $3\sigma$ error boxes
  show the positions of the stars
  according to the artificial ``observed'' traditional catalogue data,
  which were known by the hounds. For 
  clarity we only show error boxes for cases K2, B2, and P2 (see
  Table~\ref{tab1}). For details on how the catalogue data were generated
  see \S~\ref{stellarparameters}.}   
\label{fig1}} 
\end{figure}

For convenience we give a short summary of the results from the asteroFLAG
Exercise\#1 \citep{Chaplin08} 
relevant for this paper. That exercise concerned the
extraction of the large frequency spacings from simulated p-mode spectra 
of three artificial main-sequence stars designated
Katrina, Boris, and Pancho. Their locations in the H-R diagram are
shown in Figure~\ref{fig1}.
For each star, 4-year time series were generated for all 
combinations of the following parameters: apparent magnitude
$V=9,11,13,15$; inclination of rotation axis relative to the line of sight
$i=0^\circ,30^\circ,60^\circ$; and rotation equal to one, two, and three
times solar. 
\begin{figure}
\includegraphics{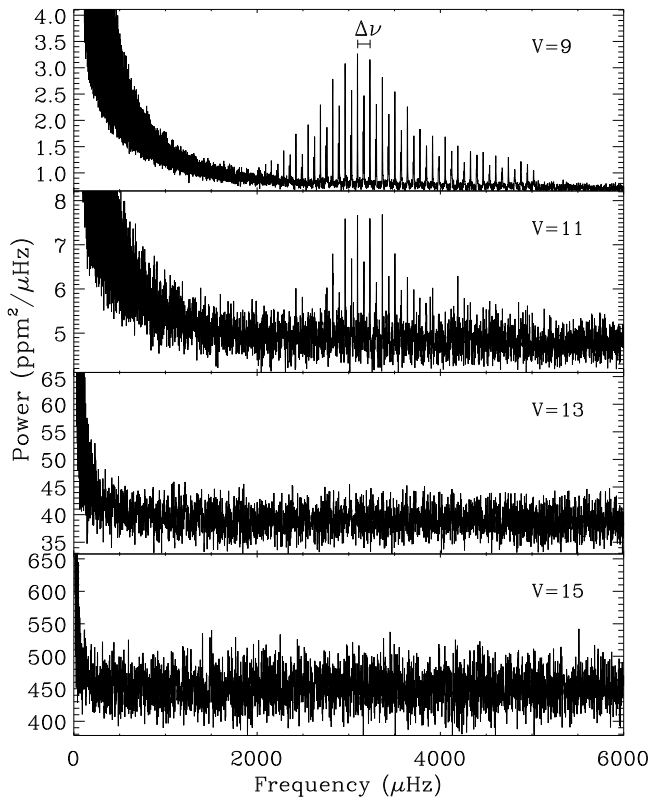}
\caption{Fourier spectra of 4-year Kepler time series of the artificial
  star Boris. Each panel corresponds to different apparent magnitudes, and
  shows co-added spectra from subdividing the time series in four day subsets.
  The large frequency spacing, \dnu, between two modes of successive
  order $n$ is indicated in the top panel.
  The catalogue data for each case are shown in Table~\ref{tab1}.
\label{fig2}} 
\end{figure}
In Figure~\ref{fig2} 
we show four examples of power spectra for Boris ($V=9,11,13,15$). While
the oscillations and the characteristic frequency spacing, \dnu, are
clearly seen for the brighter cases we can hardly see the 
excess power for the fainter ones. However, we saw from \citet{Chaplin08}
that the large frequency spacing can still be extracted  
in most cases in the tested magnitude range. The results from the exercise are
shown in the last 
column of Table~\ref{tab1}. We note that the ability to measure \dnu~for
the faintest case of Boris ($V=15$), and perhaps $V=13$, should be 
viewed with some caution. This is because the 
hounds in Exercise\#1 had access to information about the frequency range
in which they should measure \dnu~from their analysis of the brighter
cases where power excess is clearly visible. Hence, in a true blind case,
as with real data, it might be harder to establish \dnu~for such faint
cases of `Boris-like' stars. {\change For the three
faintest cases of Katrina ($V=11,13,15$) the large frequency
spacing could not be determined, preventing an asteroseismic constraint
on the radius.}
\input{tab1}

The time-series data used in Exercise\#1 were generated with the asteroFLAG
simulator (Chaplin et 
al. 2009, in preparation). Each time series included photometric
perturbations from p-modes, granulation, activity, photon noise, and
instrumental noise. 
The stellar models were generated using the Aarhus stellar evolution code,
\astec~\citep{DalsgaardAstec08} using the simple but fast EFF equation of state
\citep{Eggleton73} along with the \citet{GrevesseNoels93} solar mixture
using the OPAL tables at high temperature and the \citet{Kurucz91} table
at lower temperature. {\change The exact model values for $L$, \teff, and $R$ are
listed in Table~\ref{tab2}}. The frequencies of the p-mode signal were calculated
using the adiabatic pulsation code \adipls~\citep{DalsgaardAdipls08}. The
amplitude and damping rates were from semi analytical models \citep{Chaplin05}.
We refer to \citet{Chaplin08} for further details on the simulations and
Exercise\#1 results.



\section{Asteroseismic radius determination: Exercise\#2}\label{observations}
The aim of Exercise\#2 is to use the large frequency
spacings from Exercise\#1, together with `traditional'
stellar parameters, to estimate the radii of the stars. 
{\change While individual mode frequencies, and hence detailed asteroseismic
  analysis, are expected for the brighter
  stars, in this paper we only explore the benefit for the radius estimate
  from having access to the large frequency spacing}.
This work is an effort towards the development of robust algorithms for
estimating radii of stars of different properties and noise levels in an
automated pipeline fashion.
It is the intention that the results are used to 
indicate what radius
precision we can obtain on stars in the $V=9-15$ range, as will be
observed by Kepler. 

\subsection{Stellar parameters}\label{stellarparameters}

In addition to the large frequency spacing obtained from
asteroseismology by the first group of hounds (Exercise\#1), the second group of hounds  
(Exercise\#2 -- this paper) were given a set of
traditional catalogue data for each star.
These are given in Table~\ref{tab1}. 
Each parameter, $\log g$, $\log (Z/X)$, \teff, and $\pi$ was obtained by
adding to the   
true value (which the hounds did not know), a random number drawn from a
normal distribution having zero 
mean and standard deviation equal to the adopted uncertainty on the
parameter. 
Based on the traditional catalogue data we have estimated the locations of
the stars in the H-R diagram, which are shown in Figure~\ref{fig1}
together with the true values (star symbols). 
{\change In reality most of the atmospheric parameters, such as \teff,
$\log g$, and metallicity, will come from the Kepler Input Catalogue,
  which is based on calibrated photometry.  However, for many of the most 
  interesting targets the photometric information will be supplemented by
  high-resolution spectroscopy \citep{Latham05}. Hence, for each star at
  each magnitude we analyzed two cases. In one, the uncertainty in \teff~
  was assumed to be $200\,$K, which is representative for the relevant stars 
  from the Kepler Input Catalogue (T. M. Brown, priv. comm.). The 
  other case assumed a smaller uncertainty in \teff, which would be the
  case if additional ground-based data, such as high-resolution spectra
  \citep{Sousa08}, were available.}  

For the parallaxes we adopted Hipparcos-like precisions, with some
extrapolation at the faint end {\change to represent a conservative
  estimate of the expected astrometry from the Kepler data.} 
{\change For $V\gtrsim13$ the Hipparcos-like parallaxes had 
  large uncertainties, and we did not even generate predicted parallax data
  for $V=15$. For this faint end, the luminosity could only be estimated
  very roughly using the traditional stellar parameters.}

The large frequency spacings quoted in Table~\ref{tab1} are the averages estimated
from Exercise\#1, and the uncertainties represent the scatter of the estimates. 
These values were made over results from all hounds on all datasets at each
$V$ value. We recall that datasets in Exercise\#1 were made for different
internal rates of rotation, and different angles of inclination. 
Hence, the uncertainties given in Table~\ref{tab1} not only reflect the impact of
reduction noise due to difference in analysis between hounds but also more
subtle contributions from the different rotation and inclination.

For the faintest case of Pancho ($V=15$) there were outliers in the
individual \dnu~estimates returned by each hound in Exercise\#1, and we
decided to examine two versions of this star: one
where \dnu~was the  
mean of all the values returned by each hound,
and the other where outliers were removed before calculating the mean. 

{\change Finally, we also considered cases for all stars where the large
  frequency spacing was not used by the hounds. These are listed in the
  bottom part of Table~\ref{tab1}, below the line. This allowed us to
  compare the radius estimates with and without the asteroseismic
  constraint.}

\subsection{Analysis and results}\label{analysis}
Each hound in Exercise\#2 worked independently and used
different methods to determine the stellar radii. Details are given in
\S~\ref{stello}--\ref{trialerror}. All methods rely on matching various  
parameters (e.g. $L$, $Z$, \teff, and $\Delta\nu$) from stellar
models with the corresponding observables (e.g. $V$, $\log g$,
$\log (Z/X)$, \teff, $\pi$, and $\Delta\nu$) given in Table~\ref{tab1} to
estimate the radius.

\input{tab2}

In Table~\ref{tab2} we list the radius estimates returned by the hounds
as well as the model values.  
{\change Not all hounds examined all stars.   
 By comparing the results from the top of the table (cases K1--2, B1--8, and
  P1--10) with those at the bottom (cases K3--4, B9--16, and P11--20) it is
  quite evident that including the large frequency spacing gives a
  significant improvement in our ability to determine the radius. In most
  cases, including this asteroseismic constraint 
  reduces the uncertainty in the radius to a few percent,
  which is an improvement by a factor of 5--10. We note that in the bottom part
  of Table~\ref{tab2}, the majority of the uncertainties from Hound-2
  and a few from Hound-3, seem slightly underestimated
  compared to the deviations,
  $(R_{\mathrm{hound}}-R_{\mathrm{model}})/R_{\mathrm{model}}$, from the
  true value.}

\begin{figure*}
\plottwo{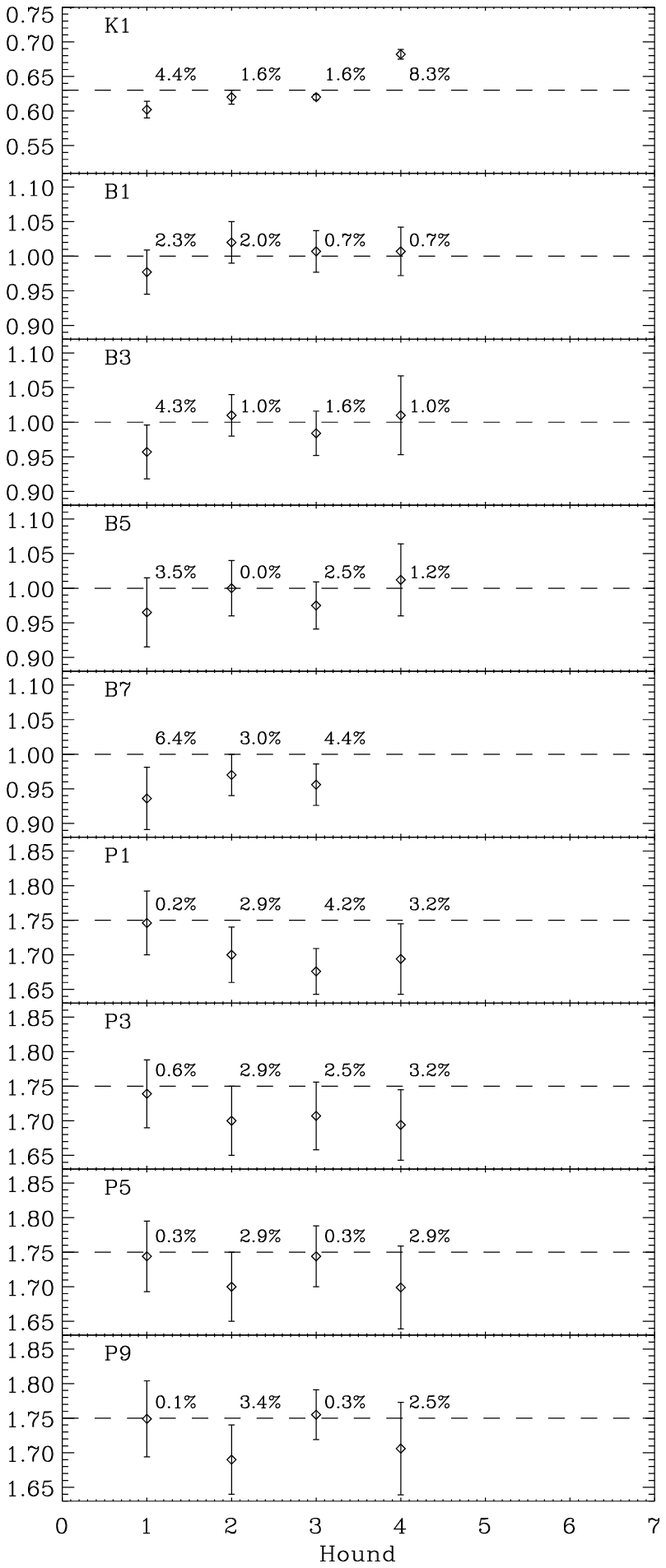}{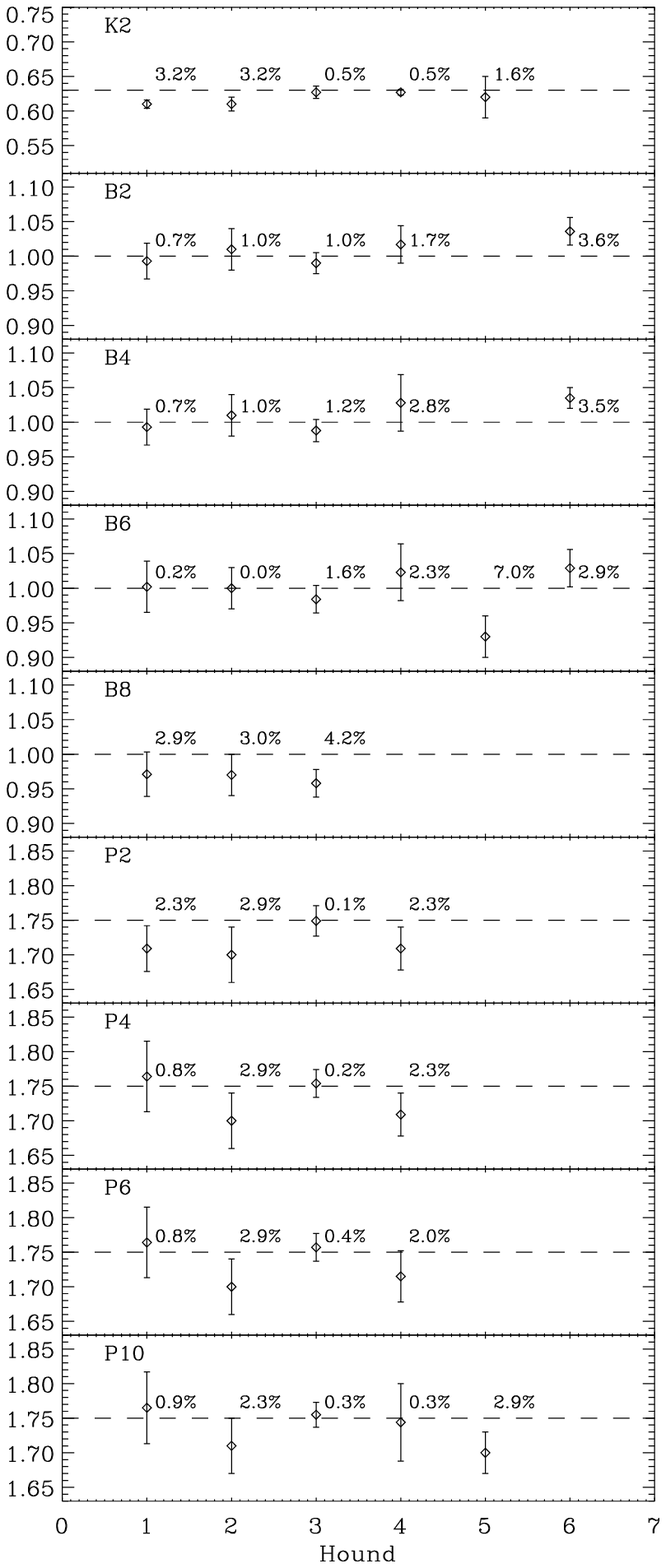}
\caption{Radii of the artificial stars and their $1\sigma$-error
  bars estimated by each hound are plotted in comparison with the true 
  value indicated by the dashed lines. We show for each case the deviation
  in percent from the true value. {\change Left panels show the results assuming
  $\sigma_{T_{\mathrm{eff}}}=200\,$K and right panels show more optimistic
  scenarios
  ($\sigma_{T_{\mathrm{eff}}}$(Katrina)$=100\,$K,
  $\sigma_{T_{\mathrm{eff}}}$(Boris)$=25\,$K, and
  $\sigma_{T_{\mathrm{eff}}}$(Pancho)$=40\,$K.)}  
\label{fig4}} 
\end{figure*}

{\change We summarize our main results in Figure~\ref{fig4}. This shows the
  results listed in Table~\ref{tab2} for 
  the stars that most hounds have examined, and which included the large
  frequency spacings. The uncertainties found by the hounds of 
  roughly 1--3\% generally seem to be consistent with the deviations that
  we see from the true values (see Fig.~\ref{fig4}). 
  We note a relatively small improvement in the radius estimates when the
  uncertainty in \teff~is significantly lower (left panels) than our base
  assumption of 200 K (right panels). This indicates that, provided we have
  the asteroseismic constraint (the large spacing), improving the estimate
  of \teff~is not particular important. 
  In the following sections we discuss in more detail the methods
  adopted by the six hounds, and refer to \S~\ref{lessons} for 
  further discussion on how our results depend on the input parameters and
  their assumed uncertainties.}

\input{section_stello}

\input{quirion_v4}
\input{bruntt_section}
\input{section_orlagh_v2}
%
\input{suarez_aflag_hh2_gteam_rev1}

\input{Boris_radius_SSousa_v3}

\section{Discussion}\label{lessons}
{\change
In this section we will discuss how our radius estimates are
constrained by our input parameters and their associated uncertainties.
We further discuss the possible systematic errors in the results.

\subsection{Dependencies on input parameters}
Figure~\ref{fig3}, which is the output from the {\sc radius} pipeline,
illustrates to some extent the effects of the various input 
parameters. All models that agree with the observations within $3\sigma$
(colored symbols) are bound by \teff, \dnu, and the metallicity.
The gravity and parallax (when available) are in general so uncertain that
they do not constrain the luminosity enough to exclude any models not
already excluded by \teff, \dnu, and the metallicity. Only in a few of the
brightest cases ($V=9$) does the parallax exclude some additional models,
reducing $\sigma_{R}$ slightly. Hence, despite the increasing
uncertainty in the luminosity for the fainter stars we basically get
the same radius and $\sigma_{R}$ (see Table~\ref{tab2}). It is in fact
mostly the increase in 
$\sigma_{\Delta\nu}$ that is responsible for the slight increase in
$\sigma_{R}$ towards fainter stars. The redundancy of $\log g$ is further
confirmed by re-running the analysis for a few of the stars with increased
uncertainty $\sigma_{\log g}=0.2$, and again we obtain the same
results within the uncertainty. 

Again, if we use Figure~\ref{fig3} as illustration we
see that $\sigma_{\Delta\nu}$ and $\sigma_{T_{\mathrm{eff}}}$ contribute
roughly equally to the uncertainty in the radius. 
However, it is evident that the largest contribution comes from the
uncertainty in the metallicity. It is $\sigma_{\log (Z/X)}$ that is
responsible for increasing the acceptable range of radius values. It will
therefore require a much lower $\sigma_{T_{\mathrm{eff}}}$ and
$\sigma_{\pi}$ to reduce $\sigma_{R}$ by the same amount as we will achieve
by a slightly better determined metallicity. 
We regard the applied
$\sigma_{\log (Z/X)}=0.1$ as realistic, but if one can reduce the
uncertainty in the observed metallicity by a factor of two to four, we can
reduce the uncertainty in the radius down to $\sim1\%$, depending on the
position of the star in the H-R diagram. 
Only in the case of a very poorly determined \dnu~does
$\sigma_{\Delta\nu}$ contribute more than $\sigma_{\log (Z/X)}$ to 
the final $\sigma_{R}$ (see Table~\ref{tab2}; cases P7 and P8). 

We further note from Figure~\ref{fig3} that the models of constant \dnu~for
a fixed metallicity, say the black squares, run almost parallel to lines of
constant radius (dashed). 
This explains why decreasing $\sigma_{T_{\mathrm{eff}}}$ has so little
effect on $\sigma_{R}$ (see Table~\ref{tab2}; cases B1, B3, B5, and B7 versus
B2, B4, B6, and B8).
This parallel alignment is even more pronounced for the hotter
star Pancho, which therefore shows no decrease in $\sigma_{R}$ as we
decrease $\sigma_{T_{\mathrm{eff}}}$. For the cooler star
Katrina, the alignment is not as good, and a better-determined
\teff~does improve the radius estimate. 
A similar illustration is shown in Figure~\ref{fig:col}, which shows that
the best fitting models (low $\chi^2$) lie along contours of constant
radius despite having a large range in age and mass. 
The tight correlation between $R$ and \dnu~seen in Figure~\ref{fig_radius}
further illustrates that this relation is not strongly dependent on the
initial physics (mass, mixing length, initial helium abundance) of the
star, which underpins that \dnu~is a powerful measure to constrain the
stellar radius. 

In summary, provided we have access to the large spacing, only metallicity
and to less extent \teff~provide additional constraints to the radius,
while the parallax and $\log g$ are redundant due to their large
uncertainties. 
}

\subsection{Systematic errors}
For a given star, the large spacing is not strictly the same at all
frequencies in the p-mode spectrum, but varies smoothly
with frequency. Hence, to correctly compare observations and models, 
it is necessary to use the same frequency range in both cases when \dnu~is
derived. However, this was not attainable in this study because the
adopted observed large spacing was the average of all values returned
by each hound of the previous Exercise\#1, each of whom used different
frequency ranges. Fortunately, it is straightforward to employ a common
frequency range for the observations and the models in the final pipeline
for analyzing real Kepler data.

We further note that all hounds used only the radial modes to derive 
\dnu~from the models, while the observed values included information on all 
detectable modes. This introduced a systematic error on the radius measurement 
because modes of different angular degree $l$ have slightly different 
spacings. For a solar model this difference is 0.5\% at most. Since
\dnu~$ \propto\rho^{0.5}$ we can get a rough estimate of the corresponding
systematic error on the radius of $<0.3$\%. However, 
this again can be corrected for and the effect be eliminated.

When fitting stellar evolution tracks to the stellar location in
the H-R diagram, a common strategy is to restrict the search to the 
1-$\sigma$ error box ($L$, \teff). Our investigation clearly shows that
this approach can lead to wrong results. With the very precise measurements
of \dnu, we saw stars where traditional parameter values (e.g. $L$,
\teff) would deviate by more than 1$\sigma$, in order to match \dnu~within
3$\sigma$, which explains why some of the stars fall outside their
1-$\sigma$ error box in Figure~\ref{fig1}. 
Statistically this is of course not
surprising, but we still find it important to stress the point.

In addition to the statistical aspect, inconsistencies can arise if there
are systematic differences between the physics included in the modeling of
the hares and the hounds or between the real star and the models.
The stellar model codes used by the hares and hounds have been developed
over many years and are well tested. For example, these
codes have been shown to give almost the same radii and luminosities, with 
differences on the order of 0.05\% \citep{Lebreton08}.
Further, comparing results from the various hounds suggest that the
detailed input physics has a relatively small effect on the radius estimate.
Recently, \citet{Kjeldsen08} suggested that the well-known offset of roughly 1\% 
between the 
observed \dnu~of the Sun and the value derived from stellar models, which
comes from improper modeling of the near-surface layers of the Sun, can be 
corrected for empirically. This offset,
which is likely to affect models of other stars as well, has not
been accounted for in this investigation because it requires knowledge of
individual mode frequencies. The systematic effect on the 
asteroseismic radius estimate from this offset is therefore expected to be
roughly 0.7\% for a Sun-like star.
 
{\change
Finally, we note that the hounds were given $\log (Z/X)$ and their
results therefore did not depend on the adopted solar value to any large
degree. However, in reality one needs to convert the observed metallicity to
$X$, $Y$, and $Z$ to compare with the models, and this conversion depends
on the assumed solar metallicity.
The range of $Z_\odot$ values currently accepted, $0.012\lesssim Z_\odot
\lesssim 0.017$  \citep{Grevesse07,Caffau08} will correspond to the same
range in $Z$ for a star similar to the Sun.  
This change, $\Delta Z=0.005$, will change the estimated radius by 
about 1--2\%. 
}

\section{Conclusions}\label{conclusions}
In an extensive hare-and-hounds exercise, we have analyzed artificial 
solar-type main-sequence stars, which are representative for targets observed
during the Kepler mission. The stellar radii found by each hound, 
using different approaches and codes for stellar evolution and
pulsation calculations, are in good agreement with one another and 
with the true values. 

We are able to determine the radius with a precision of a few percent,
and are confident that radii can be obtained with $< 3$\% accuracy for
solar-like Kepler targets on a routine basis using automatic pipeline reduction.  
We demonstrate that the reason for this is the strong relation between the
radius and the large frequency spacing, which can be measured to very high
accuracy. Our results further show that the
radius is only weakly dependent on the input physics of the stellar
models, including the equation of state and the mixing-length parameter.

The uncertainty in the radius determination is mostly dominated by the
uncertainty in the stellar metallicity, which translates to an uncertainty
in the stellar mass. In most cases the parallax and the surface gravity are
too uncertain to constrain the radius, so in essence only the uncertainty
in effective temperature and the large frequency spacing adds further to
the final uncertainty in the radius. 
{\change However, we conclude that we do not need to know \teff~to a very high
precision in order to obtain a good estimate of the radius, provided we have
access to \dnu.} 

Because the Kepler mission will provide high accuracy data for many 
binary systems, there will be cases where we will obtain both an
`asteroseismic radius', and a  `photometric radius' giving
opportunities to independently test the accuracy of our radius estimates.

While these results rely on 4-yr time series, similar results are expected
from time series of only 1--3 months but with a limiting magnitude of about
$V\sim 11$. Details on limiting magnitudes for determination of the
various frequency spacings is the aim of ongoing asteroFLAG exercises.

\acknowledgments
We are extremely grateful to the International Space Science Institute
(ISSI) for support provided by a workshop program award. This work has
also been supported by the European Helio- and Asteroseismology Network
(HELAS), a major international collaboration funded by the European
Commission's Sixth Framework Programme. DS and HB acknowledge support from the
Australian Research Council and Denison Funding from the School of Physics,
University of Sydney. SGS would like to acknowledge the support from the
Funda\c{c}\~ao para a Ci\^encia e Tecnologia (Portugal) in the form of the
grant SF RH/BD/17952/2004. SGS and MJPFGM were co-supported by grants 
{\small PTDC/CTE-AST/66181/2006} and {\small POCI/V.5/B0094/2005} from
FCT with funds from POCI2010 and FEDER. Travel support was provided by the 
AAS International Travel Grant and the High Altitude Observatory for OLC.
We acknowledge the support by STFC Science and Technology Facilities Council for YE,
WJC, STF, and RN.
 

\clearpage

\bibliography{bib_complete}



\clearpage

%
%
%

\end{document}

%% file: tab1.tex
\begin{table*}
\begin{center}
\caption{Stellar input parameters given to each hound\label{tab1}}
\begin{tabular}{llrcclrc}
\tableline\tableline
     &         & $V$   & $\log g$      &              & \teff     & $\pi$     & $\Delta\nu$ \\
Case & Star    & (mag) & (cm s$^{-2}$) & $\log (Z/X)$ & (K)       & (mas)     & (\muhz)     \\
\tableline
K1   & Katrina & 9     & 4.8(.1)       & -1.6(.1)     & 4505(200) & 46.8(1.2) & 227.82(.95) \\
K2   & Katrina & 9     & 4.8(.1)       & -1.6(.1)     & 4505(100) & 46.8(1.2) & 227.82(.95) \\
B1   & Boris   & 9     & 4.5(.1)       & -1.6(.1)     & 5780(200) & 12.5(1.2) & 135.84(.49) \\
B2   & Boris   & 9     & 4.5(.1)       & -1.6(.1)     & 5780(25)  & 12.5(1.2) & 135.84(.49) \\
B3   & Boris   & 11    & 4.5(.1)       & -1.6(.1)     & 5780(200) &  5.4(2.2) & 135.98(.59) \\
B4   & Boris   & 11    & 4.5(.1)       & -1.6(.1)     & 5780(25)  &  5.4(2.2) & 135.98(.59) \\
B5   & Boris   & 13    & 4.5(.1)       & -1.6(.1)     & 5780(200) &  3.9(3.6) & 136.89(2.81)\\
B6   & Boris   & 13    & 4.5(.1)       & -1.6(.1)     & 5780(25)  &  3.9(3.6) & 136.89(2.81)\\
B7   & Boris   & 15    & 4.5(.1)       & -1.6(.1)     & 5780(200) &           & 141.94(2.43)\\
B8   & Boris   & 15    & 4.5(.1)       & -1.6(.1)     & 5780(25)  &           & 141.94(2.43)\\
P1   & Pancho  & 9     & 4.3(.1)       & -1.4(.1)     & 6383(200) &  8.9(1.2) &  69.74(.17) \\
P2   & Pancho  & 9     & 4.3(.1)       & -1.4(.1)     & 6383(40)  &  8.9(1.2) &  69.74(.17) \\
P3   & Pancho  & 11    & 4.3(.1)       & -1.4(.1)     & 6383(200) &  4.6(2.2) &  69.74(.17) \\
P4   & Pancho  & 11    & 4.3(.1)       & -1.4(.1)     & 6383(40)  &  4.6(2.2) &  69.74(.17) \\
P5   & Pancho  & 13    & 4.3(.1)       & -1.4(.1)     & 6383(200) &  2.2(3.6) &  69.75(.16) \\
P6   & Pancho  & 13    & 4.3(.1)       & -1.4(.1)     & 6383(40)  &  2.2(3.6) &  69.75(.16) \\
P7   & Pancho  & 15    & 4.3(.1)       & -1.4(.1)     & 6383(200) &           &  69.51(10.85)\\
P8   & Pancho  & 15    & 4.3(.1)       & -1.4(.1)     & 6383(40)  &           &  69.51(10.85)\\
P9   & Pancho  & 15    & 4.3(.1)       & -1.4(.1)     & 6383(200) &           &  69.78(.33)  \\
P10  & Pancho  & 15    & 4.3(.1)       & -1.4(.1)     & 6383(40)  &           &  69.78(.33)  \\
\tableline
K3   & Katrina & 9     & 4.8(.1)       & -1.6(.1)     & 4505(200) & 46.8(1.2) &  \\
K4   & Katrina & 9     & 4.8(.1)       & -1.6(.1)     & 4505(100) & 46.8(1.2) &  \\
B9   & Boris   & 9     & 4.5(.1)       & -1.6(.1)     & 5780(200) & 12.5(1.2) &  \\
B10  & Boris   & 9     & 4.5(.1)       & -1.6(.1)     & 5780(25)  & 12.5(1.2) &  \\
B11  & Boris   & 11    & 4.5(.1)       & -1.6(.1)     & 5780(200) &  5.4(2.2) &  \\
B12  & Boris   & 11    & 4.5(.1)       & -1.6(.1)     & 5780(25)  &  5.4(2.2) &  \\
B13  & Boris   & 13    & 4.5(.1)       & -1.6(.1)     & 5780(200) &  3.9(3.6) & \\
B14  & Boris   & 13    & 4.5(.1)       & -1.6(.1)     & 5780(25)  &  3.9(3.6) & \\
B15  & Boris   & 15    & 4.5(.1)       & -1.6(.1)     & 5780(200) &           & \\
B16  & Boris   & 15    & 4.5(.1)       & -1.6(.1)     & 5780(25)  &           & \\
P11  & Pancho  & 9     & 4.3(.1)       & -1.4(.1)     & 6383(200) &  8.9(1.2) &  \\
P12  & Pancho  & 9     & 4.3(.1)       & -1.4(.1)     & 6383(40)  &  8.9(1.2) &  \\
P13  & Pancho  & 11    & 4.3(.1)       & -1.4(.1)     & 6383(200) &  4.6(2.2) &  \\
P14  & Pancho  & 11    & 4.3(.1)       & -1.4(.1)     & 6383(40)  &  4.6(2.2) &  \\
P15  & Pancho  & 13    & 4.3(.1)       & -1.4(.1)     & 6383(200) &  2.2(3.6) &  \\
P16  & Pancho  & 13    & 4.3(.1)       & -1.4(.1)     & 6383(40)  &  2.2(3.6) &  \\
P17  & Pancho  & 15    & 4.3(.1)       & -1.4(.1)     & 6383(200) &           & \\
P18  & Pancho  & 15    & 4.3(.1)       & -1.4(.1)     & 6383(40)  &           & \\
\tableline
\end{tabular}
\tablenotetext{}{Errors on each parameter are shown in parentheses.}
\end{center}
\end{table*}


%% file: tab2.tex
\begin{table*}
\begin{center}
\caption{Results from the hounds \label{tab2}}
\begin{tabular}{llcccccccccc}
\tableline\tableline
     &         & $V$   & $L_{\mathrm{model}}$ & $T_{\mathrm{model}}$ & $R_{\mathrm{model}}$ & $R_{\mathrm{hound1}}$ & $R_{\mathrm{hound2}}$ & $R_{\mathrm{hound3}}$ & $R_{\mathrm{hound4}}$ & $R_{\mathrm{hound5}}$ & $R_{\mathrm{hound6}}$ \\
Case & Star    & (mag) & $L_\odot$            & (K)                  & $R_\odot$            & $R_\odot$             & $R_\odot$              & $R_\odot$            & $R_\odot$             & $R_\odot$             \\
\tableline
K1   & Katrina & 9  & 0.15 & 4530 & 0.63 & 0.602(.012) & 0.62(.01) & 0.620(.004) & 0.682(.007) &            &             \\
K2   & Katrina & 9  & 0.15 & 4530 & 0.63 & 0.610(.006) & 0.61(.01) & 0.627(.009) & 0.627(.004) & 0.62(.03)  &             \\

B1   & Boris   & 9  & 1.00 & 5778 & 1.00 & 0.977(.032) & 1.02(.03) & 1.007(.030) & 1.007(.035) &            &             \\
B2   & Boris   & 9  & 1.00 & 5778 & 1.00 & 0.993(.026) & 1.01(.03) & 0.990(.015) & 1.017(.027) &            & 1.036(.020) \\
B3   & Boris   & 11 & 1.00 & 5778 & 1.00 & 0.957(.039) & 1.01(.03) & 0.984(.032) & 1.010(.057) &            &             \\
B4   & Boris   & 11 & 1.00 & 5778 & 1.00 & 0.993(.026) & 1.01(.03) & 0.988(.016) & 1.028(.041) &            & 1.035(.015) \\
B5   & Boris   & 13 & 1.00 & 5778 & 1.00 & 0.965(.050) & 1.00(.04) & 0.975(.034) & 1.012(.052) &            &             \\
B6   & Boris   & 13 & 1.00 & 5778 & 1.00 & 1.002(.037) & 1.00(.03) & 0.984(.020) & 1.023(.041) & 0.93(.03)  & 1.029(.027) \\
B7   & Boris   & 15 & 1.00 & 5778 & 1.00 & 0.936(.045) & 0.97(.03) & 0.956(.030) &             &            &             \\
B8   & Boris   & 15 & 1.00 & 5778 & 1.00 & 0.971(.032) & 0.97(.03) & 0.958(.020) &             &            &             \\

P1   & Pancho  & 9  & 4.53 & 6372 & 1.75 & 1.746(.046) & 1.70(.04) & 1.676(.033) & 1.694(.051) &            &             \\
P2   & Pancho  & 9  & 4.53 & 6372 & 1.75 & 1.709(.033) & 1.70(.04) & 1.749(.022) & 1.709(.031) &            &             \\
P3   & Pancho  & 11 & 4.53 & 6372 & 1.75 & 1.739(.049) & 1.70(.05) & 1.707(.049) & 1.694(.051) &            &             \\
P4   & Pancho  & 11 & 4.53 & 6372 & 1.75 & 1.764(.051) & 1.70(.04) & 1.754(.020) & 1.709(.031) &            &             \\
P5   & Pancho  & 13 & 4.53 & 6372 & 1.75 & 1.744(.051) & 1.70(.05) & 1.744(.044) & 1.699(.060) &            &             \\
P6   & Pancho  & 13 & 4.53 & 6372 & 1.75 & 1.764(.051) & 1.70(.04) & 1.757(.020) & 1.715(.037) &            &             \\
P7   & Pancho  & 15 & 4.53 & 6372 & 1.75 & 2.175(.328) &           & 1.794(.225) &             &            &             \\
P8   & Pancho  & 15 & 4.53 & 6372 & 1.75 & 2.197(.310) &           & 1.731(.264) &             & 1.70(.03)  &             \\
P9   & Pancho  & 15 & 4.53 & 6372 & 1.75 & 1.749(.055) & 1.69(.05) & 1.755(.036) & 1.706(.067) &            &             \\
P10  & Pancho  & 15 & 4.53 & 6372 & 1.75 & 1.765(.052) & 1.71(.04) & 1.755(.018) & 1.744(.056) &            &             \\
\tableline
K3   & Katrina & 9  & 0.15 & 4530 & 0.63 & 0.652(.075) & 0.59(.01) & 0.653(.022) &             &            &             \\
K4   & Katrina & 9  & 0.15 & 4530 & 0.63 & 0.652(.040) & 0.59(.01) & 0.657(.014) &             &            &             \\

B9   & Boris   & 9  & 1.00 & 5778 & 1.00 & 1.169(.140) & 1.04(.11) & 1.173(.126) &             &            &             \\
B10  & Boris   & 9  & 1.00 & 5778 & 1.00 & 1.169(.113) & 1.00(.10) & 1.152(.108) &             &            &             \\
B11  & Boris   & 11 & 1.00 & 5778 & 1.00 & 1.077(.439) & 0.94(.15) & 1.098(.281) &             &            &             \\
B12  & Boris   & 11 & 1.00 & 5778 & 1.00 & 1.077(.439) & 0.92(.09) & 1.094(.264) &             &            &             \\
B13  & Boris   & 13 & 1.00 & 5778 & 1.00 & 0.594(.550) & 0.71(.09) & 0.915(.088) &             &            &             \\
B14  & Boris   & 13 & 1.00 & 5778 & 1.00 & 0.594(.548) & 0.92(.09) & 0.931(.023) &             &            &             \\
B15  & Boris   & 15 & 1.00 & 5778 & 1.00 &             & 0.71(.09) & 1.366(.734) &             &            &             \\
B16  & Boris   & 15 & 1.00 & 5778 & 1.00 &             & 0.92(.09) & 1.383(.685) &             &            &             \\

P11  & Pancho  & 9  & 4.53 & 6372 & 1.75 & 1.299(.193) & 1.28(.15) & 1.356(.150) &             &            &             \\
P12  & Pancho  & 9  & 4.53 & 6372 & 1.75 & 1.300(.176) & 1.28(.14) & 1.336(.102) &             &            &             \\
P13  & Pancho  & 11 & 4.53 & 6372 & 1.75 & 1.001(.483) & 1.20(.18) & 1.283(.143) &             &            &             \\
P14  & Pancho  & 11 & 4.53 & 6372 & 1.75 & 1.001(.479) & 1.22(.17) & 1.297(.108) &             &            &             \\
P15  & Pancho  & 13 & 4.53 & 6372 & 1.75 & 0.833(1.36) & 1.19(.18) & 1.347(.254) &             &            &             \\
P16  & Pancho  & 13 & 4.53 & 6372 & 1.75 & 0.833(1.36) & 1.22(.17) & 1.293(.078) &             &            &             \\
P17  & Pancho  & 15 & 4.53 & 6372 & 1.75 &             & 1.19(.18) & 1.377(.502) &             &            &             \\
P18  & Pancho  & 15 & 4.53 & 6372 & 1.75 &             & 1.22(.17) & 1.294(.259) &             &            &             \\
\tableline
\end{tabular}
\end{center}
\end{table*}

%% file: section_stello.tex
\section{RADIUS: Rapid Algorithm for Diameter Identification of
  Unclassified Stars (Hound-1)}\label{stello}

In this section we describe the {\sc radius} pipeline developed and
utilized by DS to estimate stellar radii. 
The philosophy behind the development of this pipeline was to make it fast,
robust and simple.
It follows the same basic principle as the other methods described in the
following sections, namely comparing a number of observables of a
star with a grid of stellar models.
All stellar models that are within $3\sigma$ of the observations in 
all four parameters \teff, $L$, $Z$, and \dnu~simultaneously
are treated as being equally likely. 
Hence, a model either fits (is within $3\sigma$) and
is accepted or it does not fit, in which case it is discarded. 

\subsection{Models}
The selection of models (step (2) below) was based on a model grid 
that covered uniformly the entire range ($\pm3\sigma$) 
in \teff, $L$ and $Z/X$ spanned by each star (see Table~\ref{tab1}). 
The resolution in $\log (Z/X)$ was 0.1 dex, corresponding to 
$\sigma_{\log (Z/X)}$, and the resolution in mass was $0.01\,M_\odot$. 
The grid was generated with the Aarhus stellar evolution code 
\astec~using the simple but fast EFF equation of state
\citep{Eggleton73}, a fixed mixing-length parameter, $\alpha=1.8$, and an
initial hydrogen abundance of $X=0.7$. We restricted the
parameter space by fixing $\alpha$ and $X$. However, the
effect of changing these parameters was explored by other hounds (see
\S~\ref{quirion}, \ref{orlagh}, and \ref{trialerror}), who showed the
effects to be quite small.
The opacities were calculated using the solar mixture of
\citet{GrevesseNoels93} and the opacity tables of \citet{RogersIglesias95}
and \citet{Kurucz91} ($T<10\,000$\,K). 
Rotation, overshooting and diffusion were not included.
We used the adiabatic pulsation code \adipls~to
calculate \dnu~for each stellar model.

\subsection{Pipeline approach and results}
\begin{figure}
\plotone{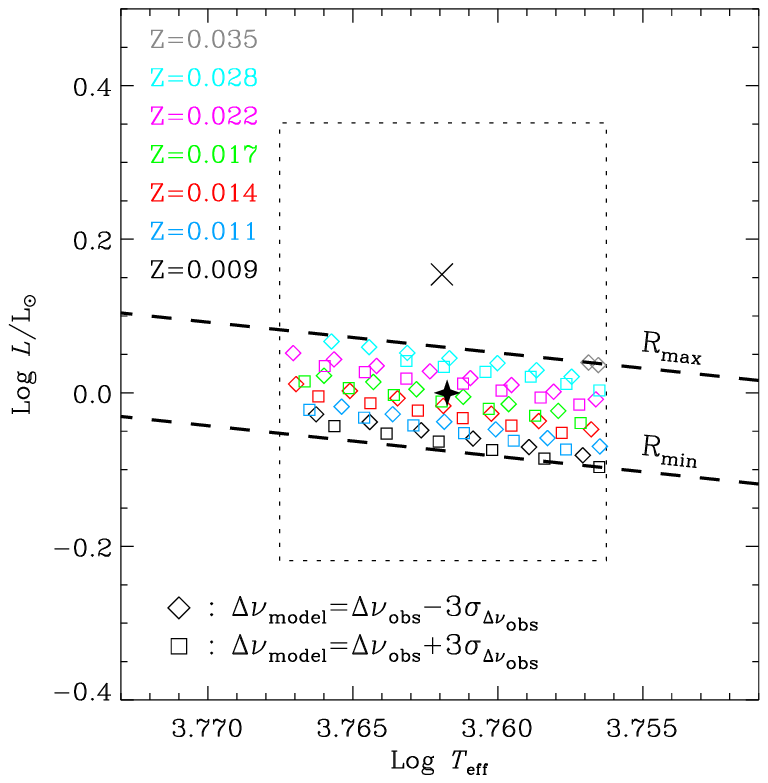}
\caption{Hound-1: Location of Boris {\change (case B2)} in the \hbox{H-R} diagram.
  The stellar location (cross) and $3\sigma$-error box (dotted) are shown
  in accord to the traditional catalogue data (see
  Table~\ref{tab1}), while the true location of Boris is shown by the star
  symbol (not known by the hound).
  Colored symbols show the location of the stellar models with large
  frequency spacings that are equal to the observed value plus (squares) or
  minus (diamonds) $3\sigma$.
  The dashed lines indicate lines of constant radius for the largest and smallest
  model that agree with observations within $3\sigma$ in all four parameters:
  $L$, $T_{\mathrm{eff}}$, $Z$, and $\Delta\nu$. Note that models with
  fixed $Z$ and \dnu~follow closely lines of constant radius.
\label{fig3}} 
\end{figure}

The pipeline 
took the following approach:  
\begin{enumerate}
 \item It determined the location in the \mbox{H-R} diagram by
 calculating $L$ (Fig.~\ref{fig3}; black cross and dotted 3-$\sigma$ error box). 
 If the parallax was not known, as in the $V=15$ cases, the
 luminosity was estimated from
 $L/$L$_\odot=(R/$R$_\odot)^2$(\teff/\teff$_\odot)^4$, where 
 $R/$R$_\odot=(g/g_\odot)/($\dnu/\dnu$_\odot)^2$.
 \item It found the stellar models that matched within $\pm3\sigma$ in $L$,
 \teff, $Z/X$, and \dnu~(see Fig.~\ref{fig3}; colored symbols).
 \item Among all matching models, the two
 with the largest and the smallest radii, $R_{\mathrm{max}}$ and
 $R_{\mathrm{min}}$, were identified. Lines of
 constant radius with these values are shown in
 Figure~\ref{fig3} (dashed lines).  
 \item The estimated radius was calculated as the average:
 $R=\langle R_{\mathrm{max}},R_{\mathrm{min}}\rangle$, with uncertainty
 $\sigma_R=(R_{\mathrm{max}}-R_{\mathrm{min}})/6$ to accommodate that we
 found the two extreme models within $\pm3\sigma$.
\end{enumerate}

In the following we give details for the two first steps: 
\begin{enumerate}
 \item[i] To calculate $L$ we interpolated the grid of bolometric corrections by
    \citet{Lejeune98}, which takes the following input: \teff, $\log g$, and
    [Fe/H]\footnote{Using bolometric corrections by \citet{Flower96}, which
      only takes \teff~as input, did not change our final results
      significantly.}. 
    For that we first converted $Z/X$ to [Fe/H] using the solar values 
    $Z_\odot=0.0188$ and $X_\odot=0.6937$ from \citet{Cox00}. We further adopted
    M$_{\mathrm{bol},\odot}=4.746$ from \citet{Lejeune98}.
    In the cases without parallax 
    we used the solar values of \teff$_\odot=5777\,$K, $\log g_\odot=4.44$
    \citep{Cox00} and \dnu$_\odot=134.92\,$\muhz~
    \citep{ToutainFroehlich92} to estimate $L$.
 \item[ii] To match models with observations we estimated \dnu~by fitting to
    the radial modes of successive order $n$ in the frequency interval 0.75--1.25 times
    \numax~(the frequency of highest power excess), 
    which we estimated from scaling the solar value (see \S~\ref{lessons} 
    for a discussion on the best choice for the frequency range). 
    The corresponding orders of $n$ for each star were:
    19--33 (Katrina), 15--27 (Boris), 13--22 (Pancho).
    The calculation of \dnu~was done for all models 
    and the results stored in the grid, making the
    search for models that matched any given star extremely fast. 
\end{enumerate}

The results from this pipeline are summarised in Table~\ref{tab2} (column: Hound-1).

%% file: quirion_v4.tex
\section{SEEK: A STATISTICAL APPROACH FOR THE AARHUS KEPLER PIPELINE (Hound-2)}
\label{quirion}
Here we describe the results obtained by a beta version of the {\sc seek}
routine developed by POQ for the Aarhus Kepler pipeline. Like {\sc radius}
(\S~\ref{stello}), {\sc seek} uses an extended 
grid of stellar models calculated with the Aarhus stellar evolution code
(\astec) and the adiabatic pulsation code \adipls~to estimate the
radius 
of stars. 

Like the {\sc radius} procedure, the {\sc seek} routine was
developed for robustness, speed and simplicity,
but we adopted a slightly more sophisticated approach.
Instead of the brute force approach of {\sc radius}, which simply
selects all models within $3\sigma$ and calculates the radius as a simple
average of the two extreme models (Fig.~\ref{fig3}),  
{\sc seek} selects the set of best-fitting models based on
a $\chi^2$ formalism and estimates the radius by fitting a
Gaussian to the distribution of radii of those models.
%

\subsection{Models}
The core of {\sc seek} is the grid of models used to fit the observations. For 
this exercise we calculated and merged 
two regularly spaced subgrids, with various values 
for the mixing-length parameter and metallicity. The first subgrid
comprised 20 sets of evolutionary tracks, each with a different combination of
the mixing-length parameter in the range $ 1.2 \le \alpha\le 3.0 $ in steps
of 0.6, and metallicity in the range $0.01\le Z\le0.03$, in
steps of 0.005. The second subgrid included 12 sets of
tracks having $ 1.5 \le \alpha\le 2.7 $, and 
$0.0125\le Z\le 0.0275$, both with the same
resolution as in the first grid. 
Each of the 32 sets comprised 57 evolution tracks  
from 0.6 to 3.0 $M_{\sun}$. The spacing between the tracks was 0.02
$M_{\sun}$ from 0.6 to 1.4 $M_{\sun}$ and 0.1 $M_{\sun}$ from 1.4 to 3.0
$M_{\sun}$.   

In this beta version of the pipeline we used simple input physics to
speed up the computation of the grid. In particular, the EFF equation of
state (Eggleton et al., 1973) was used. 
The final version will use the modern OPAL 
equation of state, covering a more extended region of the
metallicity and mixing-length parameter domains and include a range of
hydrogen fractions, which in the current version was fixed at $X = 0.7032$.
The mixture of elements was taken from \citet{GrevesseNoels93} and opacity
tables were from the OPAL package supplemented by \citet{Ferguson05} for low
temperatures. 

\subsection{Method and results}

We calculated the $\chi^2$ distribution of each star within
our grid of models using: 
\begin{equation}\label{chi2def} 
\chi^2 = \sum_{i=1}^5 \left(\frac{O_i - M_i}{\sigma_i}\right)^2,
\end{equation}
where $O_i$ are the following five observed quantities:
$\pi$, \teff, $\log g$, $\log (Z/X)$, and \dnu, each with uncertainty
$\sigma_i$ (see Table~\ref{tab1}). The corresponding model 
quantities are $M_i$. In contrast to {\sc radius}, {\sc seek} does not estimate the
observed luminosity to compare with the models, but
instead calculates a model parallax to match the observed apparent
magnitude of the star, which is assumed to be exact. 
The color transformation is done using the 
\citet{VandenbergClem03} tables.    
In the faint cases ($V=15$) we had no parallax information, hence the
parallax was not included in the $\chi^2$ sum. 
We derived the large frequency spacing for each model by fitting to
successive radial overtones of order $n\geqq15$ calculated with
\adipls.
These modes are in the asymptotic regime \citep{Tassoul80},
meaning they are close to being equally spaced, and hence
well suited for the computation of the large spacing (see \S~\ref{lessons} 
    for a discussion on the best choice for the frequency range of the modes for the \dnu~calculation).

\begin{figure}
   \centering
   \plotone{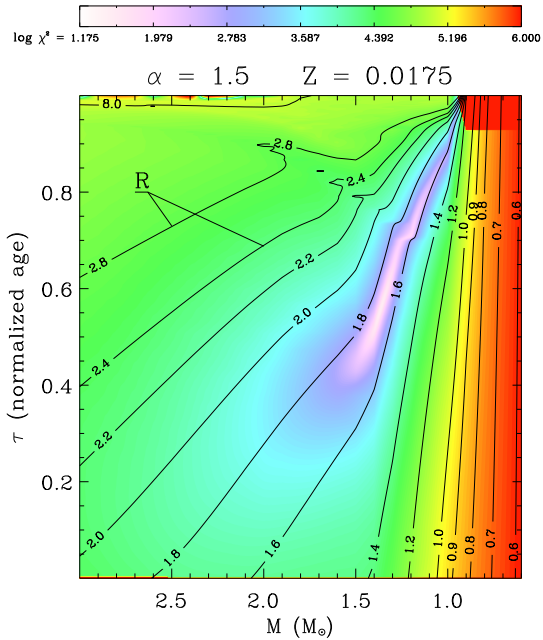}
      \caption{Hound-2: $\chi^2$ calculated for
      Pancho (case P2) in the mass-age space for models with $Z = 0.0175$
      and $\alpha = 1.5$. Radius contours are shown as solid lines (solar
      units). The color 
      follows the $\log \chi^2$ as atop of the figure. $\tau$ is the age
      normalized by the age at the giant branch, thus stars evolve parallel
      to the ordinate from the Zero Age Main Sequence at $\tau = 0$ 
      to the giant branch at $\tau = 1$. The kinks in the
      radius contours around $0.7 \lesssim \tau \lesssim 0.9$ correspond
      to the blue hook in the H-R diagram. The dark red region in the upper
      right corner marks stars that are older than 14 Gyr.}  
\label{fig:col}
\end{figure}

A typical result of the $\chi^2$ calculation is shown in
Figure~\ref{fig:col} for Pancho ($V = 9$). This illustrates a slice
of the multidimensional parameter space with a fixed metallicity, $Z =
0.0175$, and a fixed mixing-length parameter, $\alpha =1.5$. It shows 
clearly a valley of best-fitting models along a contour of
constant radius (in this case $R \sim 1.7 R_{\sun}$). 
While models with quite a range of masses and
ages fit well to the observations, they all have roughly the same
radius. 
However, due to a relatively flat bottom of the valley near the minimum
there is no single best-fitting model.
In other words the $\chi^2$ solution is degenerate. In addition, our
grid has limited resolution and the $\chi^2$-minimization problem is
highly nonlinear. Hence we cannot assume that the model with the lowest
$\chi^2$ in the grid is the absolute minimum of the problem.  Instead, we 
estimated the most likely value of the radius from a sample of best-fitting
models for which $\chi^2$ was less than a given threshold, $\chi^2_{\rm
th}$. We did this by first splitting the best-fitting models into bins of
0.005 $R_{\sun}$. For each bin we derived the total time spent by the
models at that radius, which we denoted $t(R)$. This gave us the
distribution, $t(R)$, of best-fitting radii, taking into account how
likely it was to have the observed star at each particular radius.

\begin{figure}
   \centering
   \includegraphics[angle= 90, width= 9.8 cm]{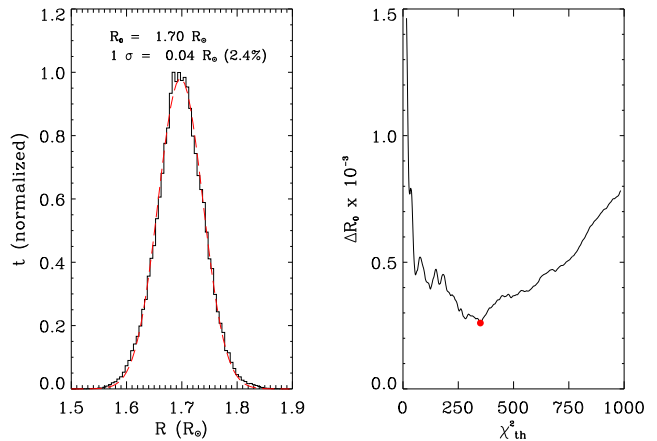}
      \caption{Hound-2: Representation of the automatic {\sc seek} process
	used to find stellar radii 
(Pancho; case P2). Left panel: Gaussian fit (red dashed line) to the
distribution histogram $t(R)$ (black solid line), of the radii of the
best-fitting models, which fulfil $\chi^2 < \chi^2_{\mathrm{th}} = 306$. The
distribution has been normalized to unity. Right panel: Fitting error on
$R_0$ as a function of the chosen threshold for selecting the best-fitting
models. The red dot shows the threshold, $\chi^2_{\mathrm{th}} = 306$ that results in
the best fit.
  }
         \label{fig:dbl}
   \end{figure}

If the threshold, $\chi^2_{\mathrm{th}}$, is well-chosen and the grid of models is
dense enough to allow a large number of models to be within that threshold,
$t(R)$ will be a normal 
distribution centred around the most likely value of the radius. We obtained
the best value for $\chi^2_{\mathrm{th}}$ by an iterative process that minimized
the difference between the median of the distribution, $t(R)$, and the
centre of a Gaussian fit. In other words, we minimized the fitting error
on the free  
parameter $R_0$, called $\Delta R_0$, by fitting the following Gaussian to
the distribution $t(R)$ (see Fig.~\ref{fig:dbl}):
\begin{equation}\label{e} 
P(R;R_{0},\sigma)= \exp({-\frac{(R-R_{0})^2}{2\sigma^2}}), 
\end{equation}
where $R_0$ is our solution for the radius. We adopted $\sigma$ as the
uncertainty on this radius, which of course was much larger than the fitting
error, $\Delta R_0$ \citep[see][ for details on the fitting
  procedure]{Press92}.  
We note that our result is not highly sensitive to the choice of
$\chi^2_{\mathrm{th}}$ as long as it is within a reasonable range
of the best value. This is because the valley is
relatively flat and has steep sides. However, if $\chi^2_{\mathrm{th}}$ is
chosen excessively big or small it will
change the width of the distribution, $t(R)$, and hence
over- or underestimate the uncertainty on the radii.

{\change
The {\sc seek} routine is able to establish the radius of the stars
without requiring an intensive computation for every new star to be
analyzed (a few seconds per star on an Intel Pentium D machine).
Our results are presented in the Hound-2 column of Table~\ref{tab2}. 

\subsection{Future development}
If a given observable is not known, the extent of 
the grid will artificially set the minimum and maximum value of this
observable. To take that into account, we will build our final grid to cover all
reasonable values in metallicity, mixing length, and mass expected for
solar-like stars. This constrained parameter space can be used to extract
information about a star for which the only well known parameter is \dnu.
If we take a star with an uncertainty of $1\,$\muhz~or less on \dnu~and only 
assume that the star is located somewhere in the parameter
space, we get a typical uncertainty of 5--10 \% on the radius. This
uncertainty is an absolute maximum if the large
spacing is known. This capacity of providing a fast and reliable answer
with minimum knowledge of the star is a powerful feature of the {\sc seek}
procedure.
}

%% file: bruntt_section.tex
\section{The Shotgun method (Hound-3)}
\label{sec:shotgun}

In this section we describe the automatic pipeline
developed by HB, called {\sc shotgun}, to perform radius estimation of stars
with or without asteroseismic constraints. 
Similar to {\sc radius} and {\sc seek}, it uses a grid of stellar models,
but differs by 
selecting a random sample of those models from which the radius is
determined. While {\sc shotgun} is still quite fast ($\sim1\,$minute per
star), it is somewhat slower than {\sc radius} and {\sc seek}. We note that
the {\sc shotgun} method is the only one presented here that does not make
use of stellar pulsation calculations, but obtains the large spacing from
scaling the solar value.

\subsection{Models}
We applied canonical scaled-solar 
BaSTI\footnote{Available from {\tt http://albione.oa-teramo.inaf.it/}} isochrones 
\citep{Pietrinferni04} in version $4.1.0$ with a mass-loss parameter $\eta=0.2$. 
The grid we used included stellar masses from $0.5$ to $5\,M_\odot$ with
steps of typically $0.02\,M_\odot$ and  
metallicities\footnote{Grid values were $Z=0.001$, $0.002$, $0.004$, $0.008$, $0.01$, $0.0198$, $0.03$, and $0.04$.} 
from $Z=0.001$ to $0.04$.
We converted the input parameter $\log (Z/X)$ to values of $Z$ by applying
the hydrogen content of the Sun, $X_\odot=0.7395$, 
in agreement with $Y_\odot=0.2485$ \citep{Basu04} and $Z_\odot=0.012$ \citep{Grevesse07}.
Each model had a set of values $({\rm age}, Z, M, L, T_{\rm
  eff})$ and for use in the following 
we calculated in addition $R/R_\odot=(T_{\rm eff}/T_{{\rm
    eff},\odot})^{-2} \, (L/L_\odot)^{0.5}$ 
and the large spacing by scaling from the Sun, 
$\Delta \nu=\Delta \nu_\odot \, (M/M_\odot)^{0.5} \, (R/R_\odot)^{-1.5}$ \citep{KjeldsenBedding95}.
For $\Delta \nu_\odot$ we used $134.8\,\mu$Hz from \citet{Kjeldsen08a}.

\subsection{Method and results}
{\change To select an unbiased sample of models from the discretely 
distributed BaSTI grid, {\sc shotgun} first generates points from a
non-discrete random distribution in the three-dimensional space $L$, \teff, and
\dnu, and finds the corresponding best matching models. }
The software is an IDL code that generates a `shot' that is comprised 
of 50 randomly chosen three-dimensional points or `pellets',
$P_i = (L_i, T_{{\rm eff,i}}, \Delta \nu_i), i\in[1;50]$, 
with mean values and 1-$\sigma$ Gaussian random errors as listed in Table~\ref{tab1}.
The value of $L_i$ is calculated from 
the $V$ magnitude and the parallax, using bolometric corrections 
from \citet{Bessell98}, which depend mostly on \teff\ and only weakly on $\log g$.
{\change When no parallax is available $L_i$ is ignored in equation
  (\ref{eq:shotgun}) below.} 
For each pellet the closest matching model in the
BaSTI isochrones is found by identifying the highest
value of the following weight function:
\begin{eqnarray}
W_i =  & \nonumber \\ 
       &  \exp \left [ - \frac{(L_i-L_B)^2   }{2\,\sigma^2_{L}   }  -
          \frac{(T_{{\rm eff},i}-T_{\rm eff,B})^2}{2\,\sigma^2_{T_{\rm eff}} }  -
          \frac{(\Delta \nu_i-\Delta \nu_{B})^2  }{2\,\sigma^2_{\Delta \nu}  }  \right ] 
\label{eq:shotgun}
 \end{eqnarray}
where $B$ are the models in BaSTI. This is similar to finding the
lowest $\chi^2$. 
\begin{figure}
\centering
\includegraphics[width=8cm]{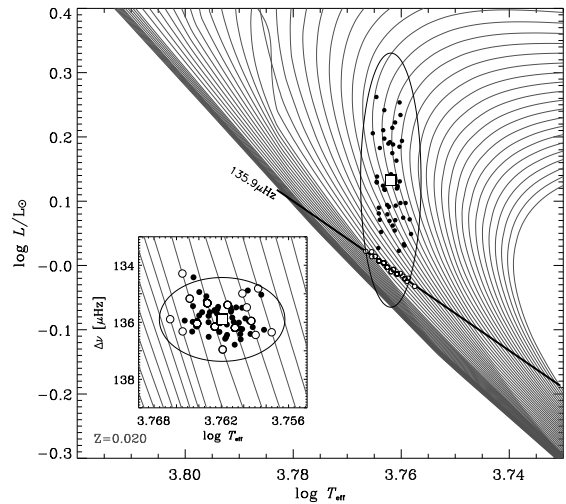}
\caption{
Hound-3: H-R diagram showing isochrones from BaSTI for $Z=0.03$. 
The square symbol marks the input value for Boris (case B2) and 
the black dots are the 50 pellets generated by
the {\sc shotgun} software. 
The open circles mark the best model match for each pellet, defined as the
highest value of the weights from Eq.~\ref{eq:shotgun}. The 3-$\sigma$ error
ellipse is indicated. The inset shows a similar diagram with \dnu~vs. \teff.  
\label{fig:shotgun}}
\end{figure}
We illustrate this in Figure~\ref{fig:shotgun}, which shows 
an H-R diagram with BaSTI isochrones for metallicity $Z=0.03$. 
The large square and 3-$\sigma$ error ellipse represent Boris (case B2).
The diagonal black line connects models 
on the BaSTI isochrones with the input large spacing of $135.9\,\mu$Hz.
The solid circles are the 50 pellets generated by {\sc shotgun}
and the open circles mark were we find the highest value of the weights,
$W_i$, for each pellet, hence representing the corresponding matched models.
To illustrate all three parameters in Eq.~\ref{eq:shotgun} we show a
similar diagram with the ordinate exchanged with the large spacing. 
It is clear that \dnu~dominates the weight function in this example
because the luminosity is very uncertain, 
which explains why all 50 pellets get matched to only 16 models within a
narrow range apparently independent of the luminosity of the  
fired pellets.
The output radius from {\sc shotgun} is
calculated as the mean value for the open circles
in the BaSTI model grid while taking into account multiple hits on the same model,
and the uncertainty is the 1-$\sigma$ RMS scatter.
One could in principle include a term in Eq.~\ref{eq:shotgun} for the metallicity, 
but the BaSTI grid is too coarse in $Z$. Instead, we did the analysis
for the two values of $Z$ in the grid that bracket the input value. 
Interpolating between the two values gave the final result for the radius, 
including the uncertainty associated with the uncertainty of $Z$.
Our results are listed in Table~\ref{tab2} (Hound-3). 

%% file: section_orlagh_v2.tex
\section{A global fitting approach (Hound-4)}
\label{orlagh}

We now proceed to describe the method used by OLC to estimate the stellar
radii of Katrina, Boris, and Pancho.
Like the {\sc seek} procedure (\S~\ref{quirion}), this approach is 
formulated as a $\chi^2$-minimization problem where 
the observations are used to find the best stellar model that minimizes
equation~(\ref{chi2def}). 
However, this approach differs by only using four observations 
in the
$\chi^2$-minimization (\dnu, \teff, $\log (Z/X)$, and $\log g$), while
the apparent magnitude, $V$,  and the parallax
measurement, $\pi$, were used exclusively to estimate the mass range of the
star.  

\subsection{Models}
As is the case for {\sc seek} and {\sc radius}, the stellar models were
derived using \astec~and \adipls.
The physics of the models included the OPAL95 opacity tables \citep{Rogers96}
the mixing-length treatment of convection \citep{BohmVitense58}, 
with a fixed convective core overshoot parameter $d_{\mathrm{OV}} = 0.25$,
but variable mixing-length parameter, $\alpha$,
nuclear energy generation rates, and the EFF 
equation of state \citep{Eggleton73}.

\subsection{Method}
Unlike the previously described methods, we did not use a pre-calculated grid
of models. Instead, the $\chi^2$ minimization was performed on one model at a
time where derivative information was used to calculate the subsequent models
that minimized the $\chi^2$, and so the only limitations in resolution 
were set by the output files of the models (e.g. five decimal places in solar
radius). It has the advantage of not being limited by the parameter 
space of a grid including the flexibility of changing the input physics of the 
models. 
The disadvantage is that it needs to be run for various
initial 
guesses to ensure a global minimum is found, with each run requiring new
models to be calculated, which is more time consuming than the grid
approach. 
 

Because we only had four observations to estimate the five stellar
model parameters (mass, age, $X$, $Z$, and $\alpha$),  
we had to fix one of the parameters, and we chose this to be $X$.
In order to also explore the parameter $X$, the minimization was repeated
while the fixed $X$ took values between 0.68 and 0.74 in steps of 0.01.

We use the Levenberg-Marquardt algorithm, along with 
singular value decomposition for the minimization \citep{Press92}.
The minimization is initiated giving each observed parameter, $O_i$, their
uncertainties, ${\sigma_i}$, and  an 
initial guess of the model parameters (mass, age, $X$, $Z$, and $\alpha$)
as input. The solution is found when $\chi^2$ has reached 0.5 or after 
four iterations, which is sufficient for a stable solution. The frequency 
range used to derive \dnu~of the model depended on the
star and the magnitude. Examples for the $V=9$ cases are Katrina
6058--7915$\,\mu$Hz, Boris 2500--4100$\,\mu$Hz, and Pancho
1400--1978$\,\mu$Hz.

To avoid local minima problems, various runs were executed using 
different initial guesses of the mass, the range of which we estimate
by deriving $M_V$ from $V$ and $\pi$ and subsequently using the relations between 
$M_V$ and mass from \citet{Allen73}. 
The scheme automatically found the stellar model that minimized $\chi^2$ for
each initial guess.
Because we had various runs of the minimization with different initial
masses ($N_M$) and different initial $X$ values ($N_X$), 
we essentially ran $N_M\times N_X$ minimizations and thus obtained this same
number of best-fitting models (see Figs~\ref{fig:masstau} and~\ref{fig:radteff}).  
We determined the radius of each of the best-fitting models 
and chose the radius values from the models 
whose $\chi^2$ is below a suitable threshold. 
These models correspond to the filled circles in Figures~\ref{fig:masstau}
and~\ref{fig:radteff}.
The quoted radius in Table~\ref{tab2} is the mean value of these, 
while the uncertainty is defined as half of the 
difference between the largest and smallest radius value.

\subsection{Results}
\begin{figure}
\centering
\plotone{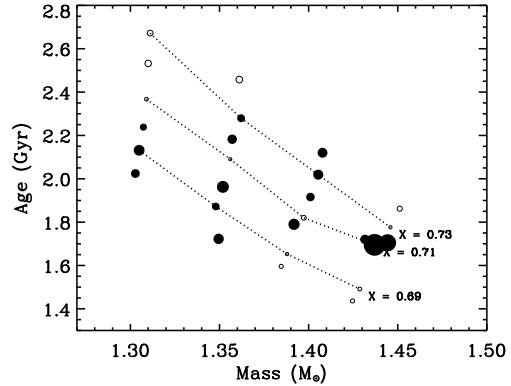}
\caption{Hound-4: Best-fitting parameters of mass and age for various minimization
  runs for Pancho (case P2). 
The larger circles represent the models with the lower $\chi^2$ values. 
The filled circles represent the models whose 
$\chi^2 < 3.9^2$. 
The dotted lines connect the results for selected values of $X$.  
\label{fig:masstau} }
\end{figure}

Figure \ref{fig:masstau} shows examples of the masses and ages of the
best-fitting models for various runs for Pancho ($V$=9).  
The initial guesses of the mass were 1.25, 1.30, 1.35, and 1.40 M$_{\odot}$,
while the $X$ value was also changed for each run. 
There are a total of 4 x 7 minimizations, and therefore we have a total
of 28 best-fitting models, each represented by a circle whose 
size is inversely proportional to the $\chi^2$ value as defined by
equation~(\ref{chi2def}) 
(i.e. a larger point means a better fit).
The filled circles are those models that give a $\chi^2 < 3.9^2$. 
For each (almost vertical) ridge in Figure~\ref{fig:masstau}, $X$ varies
between 0.68 and 0.74.
The dotted lines connects the results for $X = 0.69, 0.71$ and $0.73$.
By inspecting the figure, 
it is clear that there are correlations among the parameters, meaning that
we can obtain the same $\chi^2$ value (same dot size in
Fig.~\ref{fig:masstau}) for a range of parameter combinations.
The filled circles are those models that we accept as the best models, but
still mass, age and $X$ span large parameter ranges.
This clearly demonstrates that some of the global parameters cannot be significantly 
constrained 
by the observations that we have available in this exercise.
For example, we can see the age of the best-fitting models varies 
between a significant range of 1.4 and 2.1 Gyr, 
and the mass varies from about 1.30 to 1.45 M$_{\odot}$.
In fact, the 1-$\sigma$ uncertainties on the model with the lowest $\chi^2$ 
value are 
$\sigma(\mathrm{Mass})$ = 5\% , $\sigma(\mathrm{age})$ = 21\%, $\sigma(Z)$
= 23\% and $\sigma(\alpha)$ = 10\%. 



\begin{figure}
\centering
\plotone{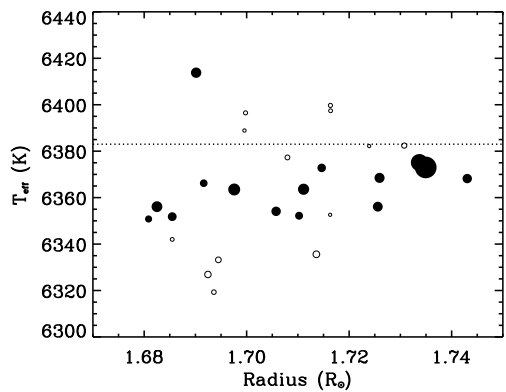}
\caption{Hound-4: Radius and effective temperature for 
the same models as in Figure~\ref{fig:masstau}.
The dotted line is the observed $T_{\rm eff}$.
\label{fig:radteff}}
\end{figure}

In Figure \ref{fig:radteff} we show the corresponding radii and effective
temperatures of the same models as shown in Figure~\ref{fig:masstau}.
The dotted line indicates the {\it observed} $T_{\rm eff}$. 
All of the models fall within  $\pm3\sigma_{T_{\rm eff}}$. 
In agreement with the results we found using {\sc seek} and {\sc radius} most
of these 
models have roughly the same radius despite spanning a large parameter range
in mass, $X$, and age. 


This automatic minimization scheme was repeated for each of the stars and 
for each $V$ value, and the results are listed in Table~\ref{tab2} (column
Hound-4).  




%% file: suarez_aflag_hh2_gteam_rev1.tex

\section{Non automated approaches}
\label{trialerror}
While the four methods described in the previous sections
(\S~\ref{stello}--\ref{orlagh}) were all automated, three relying on
large existing grids of stellar models, we now turn to methods that involve
generating small sets of models covering more limited ranges
of the parameter space, and iteratively narrowing down the parameter space
of best-fitting models manually. The fundamental idea -- fitting model
parameters to 
observations -- is still the same, but instead of being developed to the
level of fast pipeline algorithms, which can be applied routinely on large
sets of stars, the following methods again serve to demonstrate the ability
to determine precise radii.  
In addition, these methods use different stellar evolution and pulsation
codes to those in \S~\ref{stello}--\ref{orlagh}, allowing a valuable
comparison. 

\subsection{The Granada approach (Hound-5)}
\label{suarez}

In this section we present the methodology adopted by JCS, AM, and AGH
to estimate the radii of the three artificial stars (see Table~\ref{tab1}).
In contrast to the other hounds, we did not use the apparent magnitude and
the parallax information. We therefore have only one case for each star,
for which we adopted the average large spacing and the largest uncertainty
listed in Table~\ref{tab1} (case K2, B6, and P8).

\subsubsection{Modelling}

To model the stars we constructed standard 
non-rotating models from late pre-main-sequence to terminal-age main
sequence (where core hydrogen fusion ceases). We note that this does not
cover all evolution states that are consistent with the observed
parameters and 
their error bars, especially for Pancho, and our results may therefore have
systematic errors. The evolutionary  
stellar models were computed with the CESAM code \citep{Morel97}.   
Opacity tables were taken from the OPAL package \citep{IglesiasRogers96},
complemented at  
low temperatures ($T\leq10^4\,$K) by the tables by
\citet{AlexanderFerguson94}. The  
atmosphere was constructed from a grey Eddington~$T-\tau$ relation
\citep{Eddington26}.
Convection was treated with a local mixing-length model \citep{BohmVitense58} 
and we assumed various values for the mixing-length parameter
$\alpha=l/H_{\rm p}$,   
where $l$ is the mixing length and $H_{\rm p}$ is the pressure scale
height. In particular, 
values between 0.5 and 2 were explored. For the overshoot parameter 
$d_{\rm OV}=l_{\rm OV}/H_{\rm p}$ ($l_{\rm OV}$ being the penetration
length of the convective  
elements) we assumed values between 0.0 and 0.3. Such ranges of parameters
cover the range of parameters generally adopted for solar-like stars.

For the computations of adiabatic eigenfrequencies two
oscillation codes were used: FILOU
\citep[][]{TranMinhLeon95,Suarez02,SuarezGoupil08} and GraCo 
\citep[][]{Moya04,MoyaGarrido08}.

\subsubsection{Method}

{\change First, we calculated about 30 initial models that covered the observed
$\pm3\sigma$ range of each parameter, $\log g$, $\log(Z/X)$, and \teff,
while making sure they also covered various values of mixing length and
overshoot. 
From those initial models we then selected 3--5 representative models with
various masses and evolution states for which the large spacing was also
within $3\sigma$ of the observed value. The exact number of representative
models varied from star to star.
We consider the large spacing as the strongest bound on the radius. Hence,
in order to get a better radius estimate we made an additional 3--4
models around each representative model, which were then all within
$3\sigma$ of the observed $\log g$, $\log(Z/X)$, \teff, and \dnu, with the
majority within $1\sigma$ of the latter. This provided a total of about
9--20 representative models. Our final estimate of the radius
(Table~\ref{tab2}, column Hound-5) is a simple average of the radii of
those representative models ($R=\langle R_{\mathrm{model}} \rangle$) and
the corresponding uncertainty is the maximum deviation from the mean
($\sigma_{R}=\mathrm{MAX}(R_{\mathrm{model}}-\langle R_{\mathrm{model}} \rangle)$).}

%% file: Boris_radius_SSousa_v3.tex
\subsection{The CAUP approach (Hound-6)}

In this section we describe 
the procedure used by SGS and MJPFGM to estimate stellar radii.
This approach is similar to the one described in \S~\ref{suarez}, in that 
an initial set of stellar models was made based on the observed parameters
and their uncertainties. 
Then, by testing which models fit best (in the $\log g$,
$L$, \teff~parameter space), further
refinement of the range of parameter values were made by calculating
additional models. 
{\change Finally pulsation models were derived to further constrain the
parameter space using the strong bound from \dnu}.

Of the 13 stellar cases listed in Table~\ref{tab1}, the 
procedure described here was only applied on one star (Boris, case
B2, B4, and B6). The method does not rely on a pipeline algorithm,
but can be improved and used for all the stars.  
Unlike the methods described in the previous sections, this approach
only considered stellar models to be valid if they were within one sigma of
the observed $\log g$, $L$, and \teff~simultaneously. 

\subsubsection{Models}
To obtain the evolution tracks we used the CESAM code version
2K\footnote{CESAM2K available at {\tt www.oca.eu/cesam/}} 
\citep{Morel97,MorelLebreton08}.
The Livermore radiative opacities were used \citep{IglesiasRogers96}
complemented at low temperatures ($T \leq 10^4\,$K) by atomic and
molecular opacity tables from \citet{Kurucz91}. The opacities were calculated
with the solar mixture of \citet{GrevesseNoels93}, 
convection was described by the mixing length model
\citep{BohmVitense58}, 
and overshooting ($d_{\mathrm{OV}}$) was not used since this parameter is poorly known, in
particular for stars below $2\,$M$_{\odot}$ \citep{Ribas00}. The nuclear
reaction rates were from the Nuclear Astrophysics Compilation of REaction Rates
(NACRE) \citep{Angulo99}. 
For the equation of state the OPAL 2005 tables\footnote{Tables available
  at {\texttt{http://phys.llnl.gov/Research/OPAL/EOS}$\_$\texttt{2005/}}} were
used. The atmosphere of the stellar models was based on a Hopft law
\citep{Mihalas78} where convection was
not included, and radiative transfer was considered to be independent of the radiation frequency (i.e. the grey case was assumed).  
We derived frequencies of the models with the pulsation code POSC
\citep{Monteiro08}


%
%

\subsubsection{Method}

By using the parameters in Table~\ref{tab1} we located
the star and its 1-$\sigma$ error box in both the H-R and $\log g$ vs. \teff~
diagrams. 
To locate the position in the H-R diagram we estimated the 
stellar luminosity, computed from the parallax
and the apparent magnitude, using the calibration by \citet{Flower96} to
obtain the bolometric correction. 
In the calculation we
adopted the solar absolute magnitude of M$_V$ = 4.81 \citep{Bessell98}
and a bolometric correction for the Sun of 0.08 mag \citep{Flower96}.
The uncertainty on the luminosity comes mainly from the parallax.
We then built a grid of stellar models, for
different masses, $M$, initial helium content, $Y_0$, and mixing length,
$\alpha$, while having the initial metallicity ratio $Z_0/X_0$ fixed and 
equal to the observed $Z/X$. 
\begin{table*}
\begin{center}
\caption{Stellar model grids (Hound-6)\label{tab4}}
\begin{tabular}{l|cccc|ccc|cccc|ccccc|cccccc}
\tableline\tableline
Grid            & \multicolumn{4}{c}{1} & \multicolumn{3}{c}{2} & \multicolumn{4}{c}{3} & \multicolumn{5}{c}{4} & \multicolumn{6}{c}{5}           \\
\tableline 
$M/$M$_{\odot}$ & 0.94&0.97&1.03        & 1.06&1.02&1.03&1.04   &     &1.08&1.10&1.12  & 1.11&1.12&1.13&1.14&     & 1.08&1.09&1.10&    &    &     \\
$Y_{0}$         & 0.24&0.26&0.28        & 0.30&0.25&0.26&0.27   & 0.24&0.26&0.28&0.30  & 0.23&0.24&0.25&0.26&0.27 & 0.23&0.24&0.25&0.26&    &     \\
$\alpha$        & 1.20&1.40&1.60        & 1.80&1.50&1.60&1.70   & 1.20&1.40&1.60&1.80  & 1.30&1.40&1.50&1.60&     & 1.30&1.40&1.50&1.60&1.70&1.80 \\

\tableline
\end{tabular}
\end{center}
\end{table*}
In the following we use Boris (case B2) as an example to describe the
process of finding valid fitting models.

In an iterative process, we calculated three grids of models (see
Table~\ref{tab4}; grid 1, 2, and 3) to determine the range in
$M$, $Y_0$, and $\alpha$, that made the models fall within
$\pm1\sigma$ of the 
observed $\log g$, \teff, and $L$, simultaneously. In this process we made
evolution tracks with all the possible combinations of $M$, $Y_0$, and
$\alpha$ within each grid. We finally computed two more grids (4 and 5)
with higher 
resolution but covering roughly the same parameter space as the third
grid. With our adopted fixed ratio $\log (Z_0/X_0)=-1.6$, the range in $Y_0$
corresponded to $Z_0=0.017$--0.019, and $X_0=0.68$--0.75. We note that this
is a relatively small change in metallicity compared to the uncertainty
quoted in Table~\ref{tab1}.
Hence, it is possible that we 
do not see the full effect on our radius estimate from the uncertainty in
the metallicity. 

From all the grids, roughly 300 models from 32 tracks were within
$\pm1\sigma$ in $\log g$, $\log L$, and \teff~simultaneously.
From this selection of models we picked one reference model on each track, and compute
the pulsation frequencies in order to obtain the large frequency spacings,
\dnu, which we derived by fitting successive radial
overtones in the frequency range 2400--4000$\,\mu$Hz. Depending on the
models the corresponding radial orders fell in the range $n=16$--30.
To save time computing pulsations models, we estimated \dnu~for the rest of
the models along each track using the following scaling relation: 
\[
 \Delta \nu = (M/M_{\mathrm{ref}})^{1/2} (R/R_{\mathrm{ref}})^{-3/2} \Delta \nu_{\mathrm{ref}},
\]
where $\Delta \nu_{\mathrm{ref}}$, $M_{\mathrm{ref}}$, $R_{\mathrm{ref}}$
are the large frequency spacing, the mass, and the radius of the respective
reference model.

We finally selected the best two thirds of the models (188 models in the
case B2), which showed the smallest deviation between our estimated large 
spacing and the observed value.

\begin{figure}
\centering
\plotone{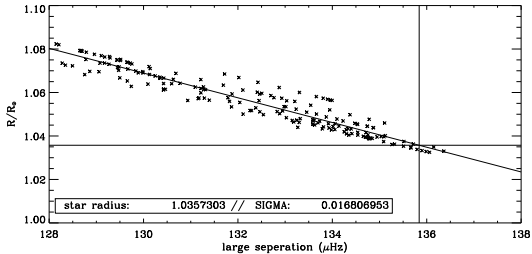}
\caption[]{Hound-6: Radius vs. large spacing for the 188 best-fitting models within one sigma
in $\log g$, $\log L$, and \teff~(Boris, case B2). {\change The vertical line
shows the measured \dnu~and the horizontal line is the corresponding radius
found by this approach.}}
\label{fig_radius}
\end{figure}

In Figure~\ref{fig_radius} we plot the relation between the large
spacing and the radius for those models. 
We made a simple linear fit to the points
presented 
in Figure~\ref{fig_radius} and used it to estimate the radius of the star,
by evaluating the value of the fit at the observed large spacing (see
vertical and horizontal lines in Figure~\ref{fig_radius}).  

To estimate the uncertainty on the radius we combined the uncertainty from  
the fit and the uncertainty in the observed large frequency spacing.
We obtained the latter by
multiplying the uncertainty on the observed large frequency spacing with the
slope of the fit. The uncertainty from the other observables are included in the
uncertainty of the fit, since the points used for the fitting were obtained
considering all models within the error boxes.
The results for the cases B2, B4, and B6 can be seen in Table~\ref{tab2}
(column Hound-6).   

%
%
